# Forward Backward Similarity Search in Knowledge Networks

Baoxu Shi[1], Lin Yang[1], and Tim Weninger[1]

[1]*Department of Computer Science and Engineering, University of Notre Dame*

**Abstract**

Similarity search is a fundamental problem in social and knowledge networks like GitHub, DBLP, Wikipedia, etc. Existing network similarity measures are limited because they only consider similarity from the perspective of the query node. However, due to the complicated topology of real-world networks, ignoring the preferences of target nodes often results in odd or unintuitive performance. In this work, we propose a dual perspective similarity metric called Forward Backward Similarity (FBS) that efficiently computes topological similarity from the perspective of both the query node and the perspective of candidate nodes. The effectiveness of our method is evaluated by traditional quantitative ranking metrics and large-scale human judgement on four large real world networks. The proposed method matches human preference and outperforms other similarity search algorithms on community overlap and link prediction. Finally, we demonstrate top-5 rankings for five famous researchers on an academic collaboration network to illustrate how our approach captures semantics more intuitively than other approaches.

Computing the similarity of two or more objects in an information network is the main focus of a large amount of scientific research and technological development. Friendship recommendation in social networks is one example, but web search, community detection, general link prediction, list augmentation, and dozens of other application areas are all singularly dependent upon some notion of similarly in the underlying networks.

Similarity is multi-faceted; various traits can be used to determine similarity depending on the specific problem domain. Entire fields of research are dedicated to the development of algorithms that effectively and efficiently retrieve objects similar to some query-object, *e.g.*, information retrieval, computer vision, and databases (broadly speaking). Researchers and practitioners understand that network topology plays a critical role in the identification of object similarity [26, 27, 35]. An appreciation of the topological features has led to the development of models of network growth, clustering, prediction, and classification.

Given a query vertex $u$, what we need is a network similarity metric that finds a target vertex $v$ to be similar if they satisfy the following criteria:

1. $u$ is highly connected to $v$, *and*

2. $v$ is highly connected to $u$

A typical approach used to compute personalized search is to measure the similarity between some query node and a set of candidate target nodes (maybe all other nodes). After the similarities of the candidate nodes have been found, the user is typically presented with a top-K list of candidate nodes ordered by their similarity scores.

For example, in citation networks Case *et al.* had previously defined six citation behaviors [5], which we simplify into two categories: a) intra-domain citations and b) cross-domain citations. Intra-domain references often include related prior work that is directly related to the referencing paper, and are the type of references that a reader would expect to see included in the experimental comparison section of the referencing paper. On the other hand, cross-domain citations often represent paradigms, platforms, and data sets that come from a separate, loosely-related area. For example, the closely related references of this paper include references to personal PageRank [11], SimRank [12] and personal SALSA [3]; while the loosely related references of this paper include references to DBLP [21], and ArnetMiner [48] datasets, or the reference to the Spark system [53] among others. A good ranking algorithm should be able to distinguish these two types of citations and give the closely related, intra-domain references a higher score than cross-domain references.

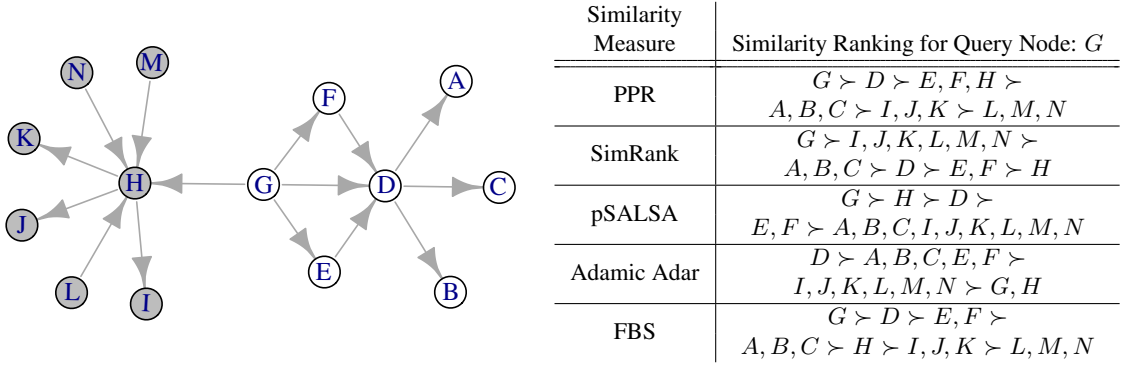

**Figure 1:** An unweighted, directed graph with two communities labeled by different colors (at left), and the vertex ranking given the query node $G$ for various graph-based similarity measures (at right).

Conventional algorithms do not work well on the citation ranking problem for a variety of reasons. To see why, consider the toy example of the citation ranking problem illustrated in Fig. 1 containing 2 communities denoted by white and grey nodes. The task is to rank the references (out-edges) with respect to the query node $G$. An ideal result would rank the intra-domain (ingroup) references higher than cross-domain (outgroup) references; even downstream references two or more links away from $G$ should, in some instances, be ranked higher than cross-domain references.

According to the forward and backward similarity criteria described above, we would expect, without loss of generality, certain properties from a ranking on the network in Fig. 1.

- The query vertex is always top-ranked, *i.e.*, $G$ is most similar to $G$ because $G = G$.

- Paper $D$ is ranked the second highest because it is directly referenced by $G$ and because other referenced papers, $E$ and $G$, reference it as well.

- Papers $E$ and $F$ are tied for third highest because they have a similar topology with respect to the query $G$; they are ranked behind $D$ because they are not referenced by other papers in the same group.

- Paper $H$ is not ranked with $D$ or $E$ and $F$ because it is does not have a large reference reciprocity, *i.e.*, $G$ does not belong to the same community as $H$.

- Papers $A$, $B$ and $C$ are tied and are ordered after $E$ and $F$ because they are directly referenced from a highly referenced paper $D$.

Further down the ideal ranking in this example we expect to find $H$ followed by its referenced papers, $I$, $J$ and $K$, further followed by incoming citations from papers $L$, $M$ and $N$.

The ranked results of many popular similarity measures including personalized PageRank (PPR), SimRank, personalized SALSA (pSALSA), Adamic Adar, and the model proposed in this paper called **Forward Backward Similarity** (FBS) is shown on the right side of Fig. 1. The ranked results clearly show that FBS, which is based on the bi-directional criteria, provides an ordering close to the ideal ordering that we expect. The differences in performance highlight the assumptions and biases inherent in the existing algorithms: 1) PPR considers $E$, $F$ and $H$ to be the same because their forward-similarities are the same from the perspective of $G$; 2) SimRank fails to assign correct similarity scores to vertices that are directly connected to the query vertex because of a problem that SimRank has with computing odd-numbered distances; 3) pSALSA is unable to distinguish indirectly connected vertices; and 4) Adamic Adar gives $A, B, C$ the same rank as $E, F$ because they all have vertex $D$ as a single common neighbor.

In general, the problematic results are due to the different interpretation of node-to-node relationships, *i.e.*, existing methods fail to consider similarity from the perspective of the candidate nodes. Although references, and directed edges in general, are one-way relations, similarity is not. Because of this oversight, the current crop of topological similarity measures may return a poor or unintuitive results.

Because network communities are often defined as being a closely connected or tight-knit groups of nodes, an inherent side effect of two-way similarity search is a greater likelihood of rating two nodes as being highly similar if they belong to the same community. So we expect that any improvement in network similarity should be reflected in the results of community detection algorithms.

The core idea of the present work is to declare two vertices $u$ and $v$ to be similar if $u$ is highly connected to $v$ *and* $v$ is highly connected to $u$.

In the following sections we present a forward backward adaptation of stochastic similarity search algorithms (FBS) that can be "plugged-in" to many existing similarity search systems. The FBS-adaptation creates a dual-perspective similarity score that satisfies the forward and backward criteria introduced above. Next, we show how the forward backward similarity search can be used to improve community analysis and link prediction, and we propose a new task called Wikipedia Category Selection that ties a given Wikipedia page with its most similar top level category. Finally, we present a qualitative study that compares the top similarity results for 5 well known data mining researchers.

# 1 Forward Backward Similarity Search

To address the problems presented in the previous section, we propose a bi-directional adaptation to stochastic search algorithms to create a forward backward similarity search (FBS) system.

Let $\mathcal{G} = (\mathcal{V}, \mathcal{E})$ denote a graph $\mathcal{G}$ containing vertices $\mathcal{V}$ and edges $\mathcal{E}$. The similarity score of $v$ given some query vertex $u$ is defined as

$$s_\mathcal{G}(u,v) = f(\pi_{\mathcal{G},u}(v), \pi_{\mathcal{G}',v}(u)),$$

where $\pi_{\mathcal{G},u}(v)$ is the similarity score of $v$ on graph $\mathcal{G}$ from the perspective of $u$, and $f$ represents an arbitrary combination function, such as linear combination

$$f(g(x), k(y)) = \lambda g(x) + (1-\lambda)k(y).$$

$\mathcal{G}'$ represents an edge reversal of $\mathcal{G}$ such that

$$\mathcal{G}' = \begin{cases} (\mathcal{V}, \mathcal{E}) & \text{for} \quad \text{undirected graph} \\ (\mathcal{V}, \{(v,u) | (u,v) \in \mathcal{E}\}) & \text{for} \quad \text{directed graph} \end{cases},$$

meaning each directed edge in $\mathcal{G}$ is flipped in $\mathcal{G}'$, and $\mathcal{G} = \mathcal{G}'$ in undirected graphs.

The similarity score $\pi$ can be calculated by asymmetric similarity metrics. In this paper we use personalized PageRank (PPR) because of its simplicity. Recall that PPR is

$$\pi_{\mathcal{G},u}(v) = \varepsilon \delta_u(v) + (1-\varepsilon) \sum_{(x,v) \in \mathcal{E}} \frac{\pi_{\mathcal{G},u}(x)}{deg_x^+},$$

where $u$ represents the query node and $v$ denotes some candidate node. $\varepsilon$ is the reset probability (a.k.a. the damping factor in PageRank), $\delta_u(v)$ is the indicator function of query node $u$ that evaluates to 1 if and only if $u = v$. $deg_x^+$ represents the out-degree of node $x$.

Simply put, this function computes personalized similarity scores from the perspectives of the query node and each candidate node. Based on the above assumptions, if two nodes are in the same community, then they are more likely to be topologically close to each other. Furthermore, by calculating personalized similarity scores from both perspectives, the algorithm should favor intra-community vertices for reasons described in the example illustrated by Fig. 1.

The dual perspective similarity metric defined here requires a combination of two separate, but related, similarity scores. Naturally, the $\pi_{\mathcal{G},u}(v)$ measure from the query nodes' perspective is called the forward similarity, and the $\pi_{\mathcal{G}',v}(u)$ similarity score from candidate nodes' perspective is called the backward similarity.

Recall that in directed graphs, like in citation networks from the examples above, a directed edge $u \to v$ represents a one-way relationship, *e.g.*, $u$ cites $v$. For example, given a query node $u$, existing similarity functions measure how important or close some candidate node $v$ is relative to all the other nodes that are cited by $u$. On the other hand, our proposed metric also measures how close or important $u$ is among all the nodes that cite $v$. The distinction here is subtle, but extremely important. Hence it is necessary to compute the backward score on the reversed graph because it preserves the network relations, and allows for the distinct forward and backward similarities to be computed.

## 1.1 Similarity Calculation

Calculating the similarity starting from a query node is rather straightforward. Indeed, all of the existing similarity methods perform this step by computing random walks or by gathering features on the network topology emanating from the query node. Computing the backward similarities from all vertices to the query vertex is less straightforward. A naive approach would calculate $\pi_{\mathcal{G}',v}(u)$ for all $v \in \mathcal{V}$, *i.e.*, the backward similarities for each of the other nodes, but this is impractical on even moderately-sized networks.

Our solution is to alternate between forward and backward search modes in order to effectively prune the backward similarity search space.

The first mode proceeds forward from the query vertex $u$, and curates a ranked list $L_n$ of the top-$n$ most similar nodes according to $\pi_{\mathcal{G},u}$, where $n \ll |\mathcal{V}|$.

At the beginning of the backward mode all edges are flipped to create $\mathcal{G}'$. From each of the top-$n$ vertices computed in the previous step, the same similarity measure $\pi_{\mathcal{G}',v}(u)$ is calculated from each node $v_1, \ldots, v_n$ in $L_n$ to the query node $u$. We only compute from the $L_n$-subset because this subset emphasizes the original rank generated from $u$'s perspective.

As a result of the forward and backward modes each candidate pair in $\{(u,v), v \in L_n\}$ has a forward similarity and a backward similarity. A single similarity measure requires some combination of the forward and backwards similarities. Any combination strategy may be used for this purpose; in the experiments section we used both a linear combination function and a saturation function.

The forward and backward mode repeats until convergence or some stopping criterion is reached. We find that repetitions slowly converge to authority-type rankings akin to PageRank scores, so its best to stop after only a few iterations.

## 1.2 Discussion

Note that in the formula, we calculate two similarity scores on two graphs, one is on the original graph from the query node's perspective and the other one is on the reversed graph from the target node's perspective.

Now that the measure has been introduced we revisit the illustration in Fig. 1. This example reveals the different roles that the forward and backward modes serve in the method. The forward mode certainly favors vertices that are near the query node $G$, especially when using random-walk based similarity measures like PPR for $\pi_{\mathcal{G},u}$. This behavior satisfies criterion 1 from earlier.

The backward mode performs the same type of similarity calculation, but begins at the one of the candidate nodes and follows inverted edges. It satisfies criterion 2 in the same way that the forward mode satisfies criterion 1. The backwards mode also produces an inherent community effect because candidate nodes $v$ that belong to the same community as $u$ are likely to have many other links from other group members that are inverted and followed outwards from the $v$ during the backward mode. Out-group vertices are more likely to be linked to from their own local communities, so their inverted edges are likely to point to other vertices in their local community instead of the query node $u$.

Using PPR as $\pi$, the example in Fig. 1 shows that $E, F, H$ have the same forward similarity score regardless of which group they belong to; $D$ receives the highest forward score because it is the most authoritative vertex relative to the query vertex $G$. By combining backward scores with forward scores, the rank of $H$ is lowered because it can only propagate $(1-\varepsilon)\pi_{\mathcal{G}',H}(H)/4$ to $G$; whereas white vertices can propagate at least $(2(1-\varepsilon)^3 + (1-\varepsilon)^2)\pi_{\mathcal{G}',v}(v)/3$ to $G$ since they are well connected to each other resulting in higher similarity scores from nodes in the ingroup.

One may ask why edges must be inverted during the backward mode. It is certainly possible to perform similarity calculation from $v$ to $u$ in the backward-mode without inverting the links, and instead use the original outgoing edges of $v$ and the other vertices. Indeed, this is exactly what happens using the commute time algorithm. However, previous surveys have found that commute time is consistently the worst performing algorithm in link prediction tasks [34]. The reason that commute time performs so poorly is because directed graphs almost always have asymmetrical edge directions. For example, Fig. 1 shows that the commute time between $G$ and every other node is undefined because $G$ has no incoming nodes, *i.e.*, the random walker can never return home.

Table 1: Graph properties of four large data sets. [†] score calculated by Pan *et al.* in 2014 [31]; [‡] score calculated by Shiokawa *et al.* in 2013 [37]; [*] value taken from Lizorkin *et al.* in 2009 [24].

| Data Set | Directed | Nodes | Edges | Avg. Degree | Bridges | Modularity | CPV |
|---|---|---|---|---|---|---|---|
| DBLP Collaboration | N | 1,464,134 | 6,249,778 | 8.54 | 0.03 | 0.73[†] | 3.93 |
| DBLP Citation | Y | 787,858 | 3,910,271 | 9.92 | 0.04 | 0.90[‡] | 1 |
| GitHub Collaboration | N | 259,977 | 1,208,438 | 9.3 | 0.06 | 0.87 | 3.39 |
| Wikipedia | Y | 15,427,669 | 123,353,354 | 5.66 | 0.02 | 0.63[*] | - |

Although the hand-drawn motivating example may be good for illustration purposes, it alone is unconvincing. So we performed a litany of experiments on several directed and undirected graphs, including ranking quality analysis, link prediction, an analysis of human preferences, and an academic collaboration case study to show the general effectiveness of the proposed similarity measure.

## 2 Experiments

The notion of relatedness or similarity plays a critical part in data mining and machine learning algorithms where models are induced by finding intra-cluster similarity and inter-cluster separability, in the case of clustering algorithms, or by drawing a hyperplane comparing class-instances in the case of classification algorithms.

Because of built-in biases and assumptions, similarity measures may succeed in one task, only to fail in many others. To show the robustness of FBS we performed four different case studies that explore the different aspects and applications of the proposed method on network data. For network clustering tasks we performed a community detection task and a co-authorship closeness case study. For network classification we performed a link prediction task and a new task that selections the Wikipedia category that best suits a Wikipedia article.

### 2.1 Data Sets

We use four real-world networks of various types and sizes. Table 1 describes the 4 networks, including their size, degree and other properties that describe the connectivity of the graph. All singleton nodes were removed.

Avg. degree is a the average number of edges (incoming and outgoing in the case of directed networks) incident to a given vertex. High average degree indicates a dense graph and vice versa. Bridges are those edges that increase the number of connected components if removed reported in Tab. 1 as the proportion of bridge-edges to total number of edges.

Modularity, a measurement of intra-cluster versus inter-cluster connectivity, is used as one estimate of the community structure of a network. It is defined in [28] as

$$Q = \frac{1}{2m} \sum_{i,j} \left[ \mathbf{A}_{ij} - \frac{k_i k_j}{2m} \right] \delta(c_i, c_j),$$

where $m$ is the number of edges in the graph, $\frac{1}{2m}$ is the normalization factor, $i,j$ are two vertices in the network, $A$ is the adjacency matrix, $k_i$ denotes the degree of $i$, and $\delta(c_i, c_j)$ is the indicator function which equals to 1 if the community of $i$ and $j$ are the same.

We performed the fast, greedy community detection algorithm by Clauset *et al.* [6] to estimate the modularity score of each network. Modularity values ranging from 0.5 and above typically indicate a distinguishable community structure, higher is better.

CPV is the community-per-vertex score, representing the average number of communities that are associated to a vertex.

**DBLP co-authorship network**: We were able to extract a co-authorship network from the January 23rd, 2015 data-dump of DBLP [21]. Vertices in this network are paper authors and two authors are connected by an undirected edge if and only if they have collaborated on a paper at least once.

**GitHub collaboration network**: This network is constructed using GitHub user activity data from `https://www.githubarchive.org` during the months of January and February 2015. Vertices in this network represent Github users. Two users are connected if they have valid contributions, *i.e.*, accepted pull requests or push operations, in the same repository at least once.

**DBLP Citation network**: The citation network is an intersection of the DBLP [21] and ArnetMiner [48] data sets. We merge all citation records from ArnetMiner into the DBLP database and remove citations connecting to papers that do not exist within DBLP. Thus nodes represent research papers and directed edges represent citations such that $u \rightarrow v$ means $u$ cites $v$.

**Wikipedia network**: This data set contains all article pages and categories in the Wikipedia data-dump from August 2014. All edges in this network are directed. An article to article edge $u \rightarrow v$ means article $u$ has a link to article $v$, article to category edge $u \rightarrow w$ indicates that article $u$ belongs to category $w$, and the category to category edge $w_1 \rightarrow w_2$ denotes that category $w_1$ is the sub-category of $w_2$. As a graph-preprocessing step, we replaced all edges pointing to redirection pages with edges that point directly to the actual target article or category page.

## 2.2 Algorithms

In the following experiments, we compared the proposed forward backward search (FBS) algorithm with personalized PageRank (PPR), personalized SALSA (pSALSA), SimRank, and the Adamic Adar index. The reset probability $\alpha$ of FBS, PPR and pSALSA is $0.15$. These iterative measures are executed until convergence with a tolerance of $10^{-6}$. The SimRank algorithm used is the fast approximation version introduced by Kusumoto *et al.* [18] with $c = 0.8$, $T = 100$ and $R = 10^4$. The Adamic Adar index measures the similarity of two vertices by taking the number of common neighbors weighted by the inverse logarithm of their respective degrees [1].

FBS requires an internal similarity algorithm for the individual forward and backward stages. We performed several tests using other similarity metrics as FBS' internal measure (not reported here) and found that PPR performed the best in nearly all cases. Therefore, PPR is used as the internal similarity measure of FBS throughout the experiments section.

## 2.3 Community Analysis

Vertices in the same community are typically considered to be more similar than vertices from different communities. To test this property, the accuracy of FBS was evaluated against ground truth communities from each dataset. Each algorithm provides an ordered list of vertices most similar to some query-vertex, based on some notion of closeness or connectedness that is typically augmented by the presence of a community-structure. A good ranking, therefore, would rank vertices that are in the same community more highly than vertices that are not in the same community. In this case, performance is measured by the overlap between the ground-truth community and the top-$k$ results.

The ground truth communities of DBLP co-authorship and citation networks were the publication venues. Repositories were used as the ground truth communities on the GitHub collaboration network. In the DBLP citation network, every node (paper) belonged to one community (venue) only, whereas in other networks nodes can be associated with multiple communities. There are no ground truth communities defined in Wikipedia. Therefore, CPV is undefined in Tab. 1 and Wikipedia results cannot be reported for this task.

The Jaccard coefficient is a standard metric for identifying the overlap of sets. In order to test overlap across rankings, we adapted the ranking metric, mean average precision (MAP), by replacing the "average precision" calculation with the Jaccard coefficient thereby creating the mean average Jaccard (MAJ) score. Specifically, the MAJ score at rank $k$ is

$$MAJ@k = \sum_{k=1}^{N} \frac{aj@k}{N},$$

where $k$ is the position in the rank and $N$ is the number of different queries. $aj@k$ denotes the average Jaccard coefficient for the top-$k$ vertices and the community to which the query-vertex belongs. Specifically, this is defined as

$$aj@k = \frac{\sum_{j=1}^{k} \sum_{i=1}^{j} \frac{|\mathcal{S}_u \cap \mathcal{S}_{v_i}|}{|\mathcal{S}_u \cup \mathcal{S}_{v_i}|}}{k},$$

where $u$ is the query-node, $v_i$ represents the $i^{\text{th}}$ ranked result, $\mathcal{S}_u$ denotes the community set of $u$, and $k$ is the number of results.

Simply put, the MAJ score measures the overlap between the ground truth and the top-$k$ algorithmic results at different levels of $k$.

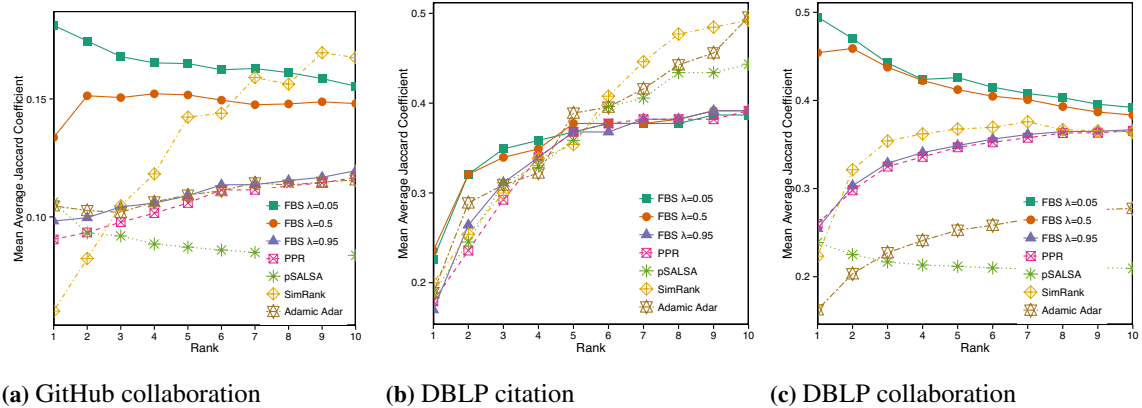

**(a)** GitHub collaboration      **(b)** DBLP citation      **(c)** DBLP collaboration

**Figure 2:** Mean Average Jaccard Coefficient score of top-10 similarity results. The higher the better. FBS outperforms all other similarity measures in the Github and DBLP collaboration datasets, and is competitive on the DBLP citation dataset.

We randomly sampled 100 query nodes, and computed the top-$k$ similarity results for $k \in [1, 10]$. To reduce the search space of FBS, we generated FBS top-$n$ list. We set $n = 20$ empirically. Among the alternatives, we used a straightforward linear combination function $f$ to combine the forward and backward modes in FBS:

$$f_{\mathcal{G}}(u,v) = \lambda \pi_{\mathcal{G},u}(v) + (1-\lambda)\pi_{\mathcal{G}',v}(u),$$

where $\lambda$ took values from 0.05, 0.5, and 0.95.

The results in Figure 2 show excellent performance for FBS with various $\lambda$ values. Even more interesting is the observation that lower $\lambda$ values tend to result in better community overlap because a lower $\lambda$ increases the importance of the backward measurements taken during the backward mode. This at least partially indicates that the backward mode significantly increases community-awareness of the similarity measure. We further find that high $\lambda$ values follows the standard PPR results, which is expected because FBS$_{\lambda=1}$ is identical to PPR.

Another interesting discovery is that local neighborhood metrics, *i.e.*, Adamic Adar, is no better than stochastic models in datasets with overlapping communities. Adamic Adar certainly favors node pairs that are within the same, single community because intra-community vertices tend to be tightly-knit and therefore have many common neighbors. Yet we find that it does not perform as well in the presence of many overlapping communities; this is probably because Adamic Adar tends to attach to the single best local community, rather than exploring the wider space of overlapping communities. This problem may be minimized when the number of communities that assigned to a node is small, *i.e.* when CPV is small as in the case with the DBLP citation network. This is probably why Adamic Adar performs very well on the DBLP citation network, but does not perform as well on the DBLP and GitHub collaboration networks which have community ratios of 3.93 and 3.39 respectively.

We were unable to include Wikipedia in these experiments as it was difficult to determine communities within the article-network. We thought to use categories as community clues, but we found that the category hierarchy is too fine grained towards the leaves, and, as well shall see in a later experiment, the categories overlap extensively in the middle and towards the top of the hierarchy.

## 2.4 Link Prediction

Link prediction has long been an important research topic, and many similarity algorithms have been developed to solve this task [22]. Typically, link prediction is performed using a supervised machine learning algorithm like logistic regression with topological and/or content features [25, 36]. In the same manner as many mainstream link prediction projects, we trained several prediction models using the similarity scores of previously named algorithms.

Specifically, we followed the experiment settings shared by Shibata *et al.* [36] and Yu *et al.* [52] that treated link prediction as classification problem. For each network, we constructed an edge set $\mathcal{E}$ containing 20,000 edges by randomly picking 10,000 true edges from the network as the positive class and randomly generating 10,000 false edges, *i.e.*, edges that do not exist in the network, as the negative class. From these 20,000 instances we performed a feature oblation case study by inducing logistic regressors trained with various feature combinations and tested each model with 5-fold cross-validation.

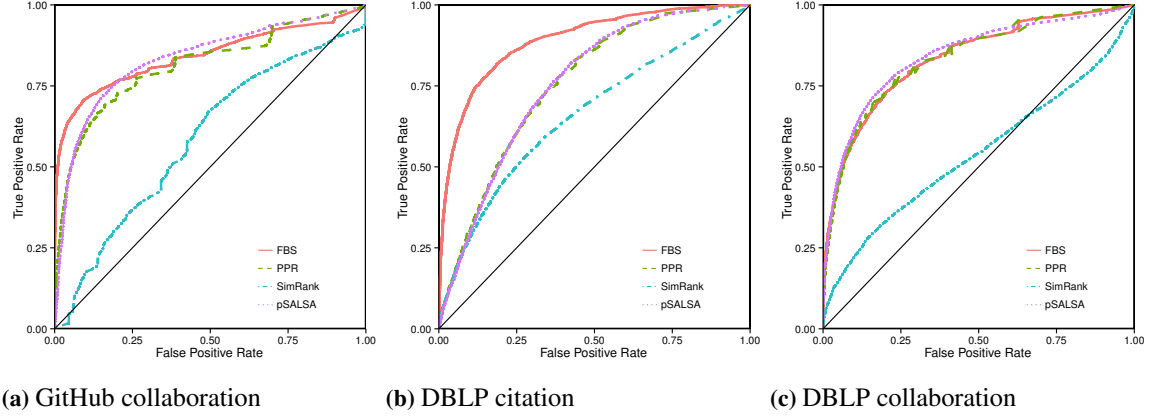

**(a)** GitHub collaboration  **(b)** DBLP citation  **(c)** DBLP collaboration

**Figure 3:** ROC curves of each link prediction model except Adamic Adar, which is shown in Figure 4. Higher is better. FBS significantly outperforms the other similarity measures on the DBLP citation graph.

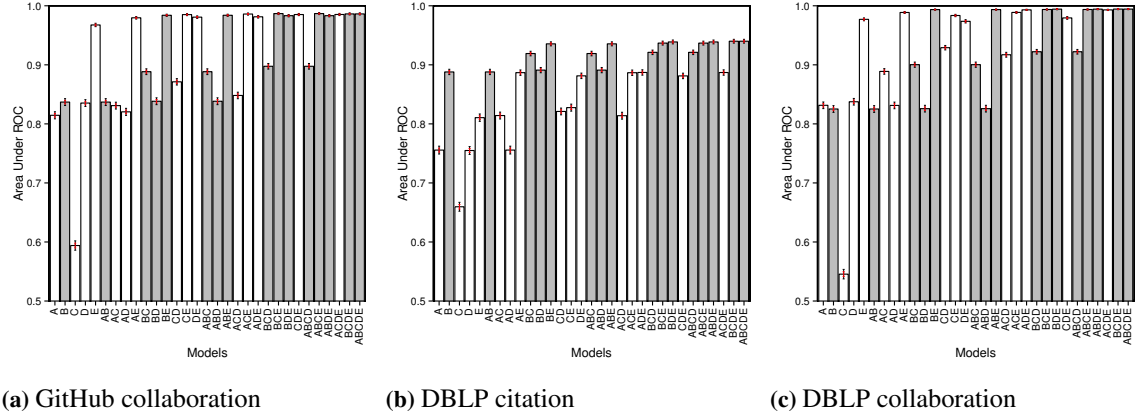

**(a)** GitHub collaboration  **(b)** DBLP citation  **(c)** DBLP collaboration

**Figure 4:** AUC scores of different link prediction models on three networks. A,B,C,D, and E represents personalized PageRank (A), the proposed FBS algorithm (B), SimRank (C), personalized SALSA (D), and Adamic Adar (E) respectively. Grey bars denote the models with FBS in its feature set. The black error bars are $95\%$ confidence intervals and the red error bars denote standard error. Higher scores are better.

Figure 3 shows the ROC curve of the logistic regression models trained with individual similarity measures; for the FBS feature we actually provide the regressor with the forward similarity and the backward similarity as two separate features, which essentially allowed the regressor pick the $\lambda$ values as the regression function parameters.

In the link prediction results of Figure 3 and 4, we find that FBS has the best AUC score on the DBLP citation network, $17.5\%$ better than PPR and $9.6\%$ better than Adamic Adar. On GitHub and DBLP collaboration networks, Adamic Adar works better than the proposed method. This is expected because in the GitHub and DBLP collaboration networks nodes that collaborate on the same paper or repository are directly connected, therefore any paper or repository can be seen as a clique in the network, heavily favoring local neighborhood measures like Adamic Adar. However, in the DBLP citation network such tight-knit communities do not exist resulting in a drop in performance for local neighborhood metrics.

Although above discussion states that DBLP and GitHub collaboration networks have denser intra-community connectivity, they actually have a lower modularity score due to their high community-per-vertex (CPV) ratio. When CPV is high, nodes in the network often belong to multiple communities, which will increase the number of inter-community edges and make the community structure less prominent. Both DBLP and GitHub collaboration networks have high CPV ratios, resulting in a less distinct community structure and low modularity score. This observation matches our expectation that FBS will perform better on graphs with more distinct community structure as indicated by the modularity score.

The results of the feature oblation tests are shown in Figure 4. The FBS measure (labeled B) appears in many of the highest scoring feature sets, but is found to underperform Adamic Adar on this task for the same reasons as above.

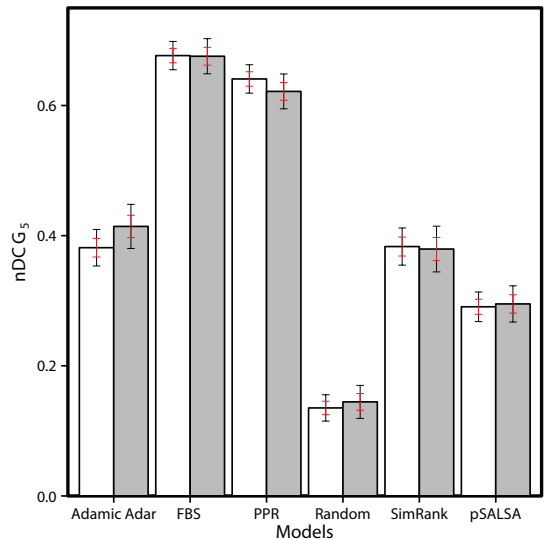

**Figure 5:** Normalized discounted cumulative gain of top-5 rankings. White bars represent the result of the original 271 random articles. Grey bars denote the result of a 182-sample subset of the original samples where the backward-mode weight is not zero. FBS is $8.7\%$ better than PPR on the subset. The black error bars are $95\%$ confidence intervals and the red error bars denote standard error. Higher values are better.

## 2.5 Wikipedia Category Selection

Apart from analyzing the performance on the community and link prediction tasks, we also performed a large scale human judgement test on a Wikipedia category selection task. In Wikipedia, each article has one or more (usually many) categories associated with it. Each category has a parent category situated in a hierarchy up to a set of approximately 40 top level categories, *e.g.*, Sports, People, Education. This category hierarchy is messy and highly overlapping; articles frequently have paths that lead to 10 or more top level categories. The task in this case study is to match each article with the best top level category.

This particular task is an effective way to judge network similarity because a "good" selection of a top level category, and especially the subcategories that lead to it, ought to reflect strong connectedness within the category hierarchy of Wikipedia.

In this task, we randomly picked 271 articles from Wikipedia and selected 10 category candidates for each article based on the union of the results of all five metrics. For each of the 271 articles, we asked 20 mechanical turk users (turkers) to "pick the Category that best fits the Wikipedia entry" from ten candidate-categories. The order of the candidate categories was randomly generated. We also sporadically inserted "gold" questions with had clear answers that the turker had to answer correctly 80% of the time in order to have their answers count and be paid[1].

In order to evaluate each similarity metric against the human preferences, we adapted the normalized discounted cumulative gain (nDCG) ranking metric from information retrieval literature. NDCG requires a ranked list of relevance scores $rel_{u,c}$, where $c$ is the candidate top level category of article $u$, which we calculate to be equal to the number of people who picked $c$ as the best top level category for article $u$. For example, if category A was picked 1 time, and category B was picked 7 times then the NDCG-relevance of A is 1 and the NDCG-relevance of B is 7. Hence $rel_{u,c} \in [0, 20]$ and $\sum_{i=1}^{20} rel_{u,c_i} = 20$.

With the relevance score of all the categories, we evaluated the top-5 most similar top level categories as calculated by each algorithm using nDCG

$$nDCG_k = \frac{\sum_{i=1}^{k} \frac{2^{rel_{u,c_i}} - 1}{log_2(i+1)}}{IDCG_k},$$

where $k$ is the cut-off position in the rank ($k = 5$ in this study), $IDCG_k$ is the ideal DCG score representing the DCG of a perfect top-$k$ ordering. For example, if $k = 3$ and the relevance scores of retrieved vertices are $\{A : 4, B : 10, C : 6\}$, then the ideal top-3 ranking is $B \succ C \succ A$ resulting in an $IDCG_3 = 1070.249$.

---
[1]Although the HIT was available on Amazon Mechanical Turk, we developed and ran the system using Crowdflower.

Table 2: Top-5 results of FBS, PPR, and Adamic Adar on DBLP co-authorship network. $\lambda = 0.5$. $^\dagger$ denotes graduate student, $^\ddagger$ represents assistant professor, $^\star$ indicates tenured professor, $^\circ$ is industry researcher, $\sqrt{}$ means advisee of the query scholar.

| | **Forward Backward Similarity** | | | | |
|---|---|---|---|---|---|
| Query | Philip S. Yu | Christos Faloutsos | Jon M. Kleinberg | Rajeev Motwani | Jiawei Han |
| 1 | Xiangnan Kong $^\ddagger\sqrt{}$ | Danai Koutra $^\dagger\sqrt{}$ | David A. Easley $^\star$ | Mayur Datar $^\circ\sqrt{}$ | Tim Weninger $^\ddagger\sqrt{}$ |
| 2 | Joel L. Wolf $^\circ$ | B. Aditya Prakash $^\ddagger\sqrt{}$ | Lars Backstrom $^\circ\sqrt{}$ | Dilys Thomas $^\circ\sqrt{}$ | Yizhou Sun $^\ddagger\sqrt{}$ |
| 3 | Ming-Syan Chen $^\star$ | Duen Horng Chau $^\ddagger\sqrt{}$ | Lillian Lee $^\star$ | Arvind Arasu $^\circ$ | Chi Wang $^\circ\sqrt{}$ |
| 4 | Wen-Chih Peng $^\star$ | Agma J. M. Traina $^\star$ | Eva Tardos $^\star$ | Tomas Feder $\sqrt{}$ | Xiao Yu $^\circ\sqrt{}$ |
| 5 | Vincent S. Tseng $^\star$ | Caetano Traina Jr. $^\star$ | David Kempe $^\star\sqrt{}$ | Shivnath Babu $^\star$ | Xifeng Yan $^\star\sqrt{}$ |
| | **Personalized PageRank** | | | | |
| Query | Philip S. Yu | Christos Faloutsos | Jon M. Kleinberg | Rajeev Motwani | Jiawei Han |
| 1 | Ming-Syan Chen $^\star$ | Caetano Traina Jr. $^\star$ | Eva Tardos $^\star$ | Jennifer Widom $^\star$ | Chi Wang $^\circ\sqrt{}$ |
| 2 | Wei Fan $^\circ$ | Agma J. M. Traina $^\star$ | Ravi Kumar $^\circ$ | Hector Garcia-Molina $^\star$ | Yizhou Sun $^\ddagger\sqrt{}$ |
| 3 | Vincent S. Tseng $^\star$ | Duen Horng Chau $^\ddagger\sqrt{}$ | Lillian Lee $^\star$ | Dilys Thomas $^\circ\sqrt{}$ | Xiao Yu $^\circ\sqrt{}$ |
| 4 | Wen-Chih Peng $^\star$ | Danai Koutra $^\dagger\sqrt{}$ | David A. Easley $^\star$ | Tomas Feder $\sqrt{}$ | Xifeng Yan $^\star\sqrt{}$ |
| 5 | Bing Liu $^\star$ | B. Aditya Prakash $^\ddagger\sqrt{}$ | David Kempe $^\star\sqrt{}$ | Mayur Datar $^\circ\sqrt{}$ | Tim Weninger $^\ddagger\sqrt{}$ |
| | **Adamic Adar** | | | | |
| Query | Philip S. Yu | Christos Faloutsos | Jon M. Kleinberg | Rajeev Motwani | Jiawei Han |
| 1 | Charu C. Aggarwal $^\circ$ | Danai Koutra $^\dagger\sqrt{}$ | Ravi Kumar $^\circ$ | Jennifer Widom $^\star$ | Yizhou Sun $^\ddagger\sqrt{}$ |
| 2 | Haixun Wang $^\circ$ | Duen Horng Chau $^\ddagger\sqrt{}$ | Prabhakar Raghavan $^\circ$ | Rina Panigrahy $^\circ\sqrt{}$ | Chi Wang $^\circ\sqrt{}$ |
| 3 | Jiawei Han $^\star$ | David A. Bader $^\star$ | Andrew Tomkins $^\circ$ | Dilys Thomas $^\circ\sqrt{}$ | Xifeng Yan $^\star\sqrt{}$ |
| 4 | Wei Fan $^\circ$ | Hanghang Tong $^\ddagger\sqrt{}$ | Eva Tardos $^\star$ | Tomas Feder $\sqrt{}$ | ChengXiang Zhai $^\star$ |
| 5 | Ming-Syan Chen $^\star$ | Evangelos E. Papalexakis $^\dagger\sqrt{}$ | David Kempe $^\star\sqrt{}$ | Hector Garcia-Molina $^\star$ | Jian Pei $^\star\sqrt{}$ |

We used the saturation combination function as $f$ for this task rather than the linear function used in the community detection and link prediction tasks earlier:

$$f_{\mathcal{G}}(u,v) = \lambda \frac{\pi_{\mathcal{G},u}(v)}{\pi_{\mathcal{G},u}(v) + k_1} + (1-\lambda)\frac{\pi_{\mathcal{G}',v}(u)}{\pi_{\mathcal{G}',v}(u) + k_2},$$

where $\lambda = 0.571$, $k_1 = 0.72$, $k_2 = 0.3$ empirically. The saturation function was used here rather than the linear function used in the community detection and link prediction tasks in order to demonstrate the flexibility of FBS and because the saturation function performed slightly better than the linear function.

The results of Figure 5 (grey bars) shows that the nDCG score of FBS is $8.7\%$ better than PPR at predicting the human annotated Category for a given Wikipedia article.

There are some confounding issues that arise in the large and highly connected Wikipedia network. We find that nearly 1/3 of the 271 randomly chosen query-articles are so far away from the query articles that they do not return any weight in the backward mode. This is because the PPR during in the forward mode tends toward 0 the as the distance and spread of the initial weight increases. The query articles and top level categories were simply too far apart, and the weight dispersion was too great for the similarity scores to propagate backwards to the query article. Thus, in 1/3 of the cases the FBS results were identical to the PPR results.

If we remove these cases, we are left with 182 query articles. These are represented in the grey bars in Figure 5. In the reconfigured experiments, the FBS has a statistically significant improvement over the PPR score. Another potential solution to this problem is reduce the reset probability from the default of $0.15$ to $0.10$ or $0.05$, etc. An experimental evaluation of the damping factor is outside the scope of this work because any stochastic similarity measure can be used within FBS in place of PPR; so we leave this study as a matter for future work.

## 2.6 Co-authorship Case Study

In order to visually illustrate the type of result the FBS measure generates, we performed a small scale case study on the DBLP co-authorship network. In this test, we generated the top-5 results for five prolific computer scientists and display the ordered results in Table 2. For comparison-sake, the results of PPR and Adamic Adar are also listed.

In this case study we revert to the simple linear combination function with $\lambda = 0.5$, meaning we measure the similarity score from both perspectives equally. In simple terms, this set up finds the nodes $v$ that are important to the query node $u$ such that the query node $u$ is also important from the perspective of $v$. In this academic collaboration case, FBS generates the rank by measuring *among all the people that the query scientist values, whose research collaboration network depends most on the query scientist?*.

A quick examination of Table 2 describes the differences in this case. In the result of PPR, the mostly highly ranked people are typically tenured professors or other highly prolific scholars. As for Adamic Adar, scholars that actively collaborate with the query scholars and their group are highly productive. Whereas in the result of FBS, the most highly ranked people are usually students of the query computer scientist due to a high collaboration community effect, where, in this case, the community of the top ranked individuals are group members or colleagues at the same university, etc.

The top-ranked person in the PPR and Adamic Adar results include, Ming-Syan Chen, Caetano Traina Jr., Eva Tardos, Charu C. Aggarwal, Ravi Kumar, and Jennifer Widom who are all established and frequent authors. But these scholars may not be as tightly connected to the query researcher because their collaboration networks do not necessarily depend on the query researcher. Whereas in FBS's result, the highly similar researchers, such as, Xiangnan Kong, Danai Koutra and Tim Weninger, are recently graduated students or current students who are still in the process of establishing their own independent collaboration networks. Simply put, in collaboration networks FBS tends to find co-dependent collaborators like advisor-student relationships, whereas PPR and Adamic/Adar methods tends to suggest more established, *i.e.*, more connected researchers.

We predict that if we run FBS on a temporal academic collaboration network with the same query vertex, we would find that graduate students at the very beginning of their career have a very low rank because of the limited number of collaborations (papers) starting off. But, as they publish with their adviser and near graduation, then they should move up in the FBS rankings. After graduation, their rank will drop as they develop their own independent collaborations or become inactive. Of course, the advisor/advisee relationship is only one of many types of collaborations in academic networks, individual rankings could be conflated and complicated in many different and interesting ways.

## 3 Related Work

Local neighborhood similarity measures (see [33] for a comprehensive study) count the number of common neighbors between two vertices weighted by the total number of edges for each vertex [1]. These local measures perform impressively on link prediction or concept similarity tasks [32]. Yet, because local similarity measures only look at the ego networks of the query and target nodes, they will not work if the query and target are separated by more than one hop, even if they are highly connected by intermediaries [47]. As a result, local similarity measures may perform poorly if they are topologically separated, *i.e.*, have no mutual friends. In contrast to local neighborhood-based similarity measures, global similarity measures are largely based on the notion of randomly walking through the network. Because global searchers are not confined to some local neighborhood, they are able to consider vertices that are more than a single hop away from the query node. The class of global network ranking measures, typified by PageRank [30], HITS [16] and SALSA [20], are excellent at computing the global importance of objects according to their overall connectivity. Numerous additional studies stem from these initial studies. Gyongyi *et al.* proposed TrustRank, a variant of PageRank with biased score propagation [9]. Wu *et al.* augmented TrustRank by introducing topical scores [49]. Sydow and Fagin *et al.* added back-step to PageRank [46, 7]. Halu *et al.* applied PageRank to multiplex networks [10]. However, these existing models do not allow for a local ranking relative to some specific query node.

**PageRank Family**  PageRank calculates the authority-scores of all vertices using the random surfer model [30], which is based on the random walker model. In the random walker model, a hypothetical walker-agent starts at some node $u$ and moves to a neighbor of $u$ at random. Starting from $u$, the expected number of steps required for the random walker to reach $v$ is called the hitting time $H_{u,v}$. In almost all cases, the hitting time is not symmetric, *i.e.*, the hitting time between $u$ and $v$ is not the same as the hitting time between $v$ and $u$. A natural solution to this asymmetry is to consider the commute time $C_{u,v} = H_{u,v} + H_{v,u}$.

One problem with these measures is their sensitivity to parts of the graph that are far away from $u$ and $v$, even if $u$ and $v$ are near each other. Personalized PageRank (PPR) [11] and similar models [9, 49, 46], correct this problem by resetting the random walker back to $u$ with fixed probability $\alpha$ at each step.

In all of these models the random walker assigns similarity scores to vertices by stochastically following *outward* links moving away from the query vertex, *i.e.*, moving forward. Under no circumstance does the walker move backwards along a directed edge, therefore we label algorithms in the PageRank family to be *forward-only* models.

**SALSA Family**  SALSA creates a bipartite graph where the left-hand-size contains vertices with more outgoing links than incoming, called candidate hubs, and the right-hand-size contains vertices with more incoming links than outgoing, called candidate authorities [20]. Each side in the bipartite graph is iteratively updated by gathering scores from the opposing side. In order to personalize the SALSA algorithm, Bahmani, *et al.* added a restart probability into the hub score calculation of SALSA [3] similar to how personal PageRank adds a restart probability to the PageRank model. Researchers at Twitter developed a similar measure called Who To Follow (WTF) that uses a bipartite graph where the left hand side is constructed form a personalized "circle-of-trust" and the right hand side is created by the those Twitter users that the query vertex follows [8]. WTF then performs SALSA-like iterations to calculate whom on the left-hand-side should be recommended for the user, *i.e.*, the query node, to follow.

**SimRank Family** Apart from personalized extensions of PageRank and SALSA, SimRank is also used for similarity search on networks based on the notion that "objects are similar if they are referenced by similar objects" [12]. P-Rank generalizes SimRank by computing structural similarity based on in-neighbors only, on out-neighbors only, and then by making a linear combination of the two similarities [54].

An important property of SimRank, P-Rank, and related measures [19, 18] are that they are symmetric, that is, the similarity between two nodes $u$ and $v$ is the same as the similarity of $v$ and $u$. Symmetry is an important quality in the notion of similarity, however SimRank and its derivatives have a severe flaw due to the symmetry requirement that causes it to fail when the path-distance between the query vertex and target is odd [15].

Each of the network similarity families have their own particular strengths and weaknesses. The PageRank family, in particular, has been successful in many applications, *e.g.*, Web search, and has been personalized (PPR) to find nodes topologically similar to some query node [11]. The SALSA family has shown some success in personalization, although the calculation of hubs and authorities scores is not naturally applicable to the computation of personalized similarity scores, and therefore requires a re-appropriation of its original intent in personalized cases.

Other methods include SCAN [50], ObjectRank [4] and PopRank [29], among many others [17]. However, these similarity measures disregard the different roles and types of the objects and links. Adoption of these homogeneous measures to heterogeneous networks has significant drawbacks: 1) objects of different roles and types carry different semantic meanings, and 2) it does not make sense to mix different types without distinguishing their semantics.

In work on heterogeneous information network analysis, similarity search was studied by distinguishing the different types of nodes in a network. By considering different linkage paths across different types of nodes in a network, advanced similarity semantics were derived, and the concept of meta-path-based similarity (PathSim) was introduced [41]. Other applications have been developed to determine the similarity between objects in a heterogeneous information network including: ranking-based clustering [45, 42, 43], ranking-based classification [13, 14], meta-path-based similarity search [41], relationship prediction [39, 40], relation strength learning [43, 38], concept similarity [23, 2, 51] and network evolution [44]. The works have resulted in the development of very powerful tools that manipulate and mine data, but, at their core lies a basic notion of network similarity based on a forward-only search.

## 4 Conclusions

In the present work we argue that network similarity should be considered from the perspective of the target *and* the source nodes. To that end, we have proposed a dual perspective similarity metric called Forward Backward Similarity (FBS) that calculates network similarity based on the perspective of both the query node and the candidate endpoint. Additionally, FBS can be "plugged-in" to many existing network similarity algorithms, thereby extending its use to many different situations.

Experiments showed the effectiveness of the proposed method in multiple scenarios. The community overlap experiment showed that FBS successfully identifies node pairs in overlapping communities where nodes are frequently assigned to multiple communities. Next we adapted FBS for use in the canonical link prediction problem; we found that the proposed algorithm works well on highly-modular networks, *e.g.*, DBLP citation network where the AUC of FBS is $17.5\%$ better than PPR and $9.6\%$ better than Adamic Adar on DBLP citation network. The results of category selection on Wikipedia showed that FBS is $8.7\%$ better than other algorithms, indicating FBS's ability to work on complicated real world networks. Lastly we showed the results of a qualitative case study on DBLP academic collaboration network with top-5 similarity search results that illustrated the different semantics FBS captured than others.

## 5 Acknowledgements

This work is supported by the Templeton Foundation under grant FP053369-M/O.